\documentclass[letterpaper]{article} %
\usepackage{aaai24}  %
\usepackage{times}  %
\usepackage{helvet}  %
\usepackage{courier}  %
\usepackage[hyphens]{url}  %
\usepackage{graphicx} %
\usepackage{natbib}  %
\usepackage{caption} %
\usepackage{algorithm}
\usepackage{algorithmic}
\usepackage{newfloat}
\usepackage{listings}
\DeclareCaptionStyle{ruled}{labelfont=normalfont,labelsep=colon,strut=off} %
\floatstyle{ruled}
\newfloat{listing}{tb}{lst}{}
\floatname{listing}{Listing}
\newcommand{\mypar}[1]{\paragraph{#1}}
\newcommand{\ie}{\textit{i}.\textit{e}.\ }

\title{MIDDAG: Where Does Our News Go? Investigating Information Diffusion via Community-Level Information Pathways}
\author{
    Mingyu Derek Ma$^{1}$,
    Alexander K. Taylor$^{1}$,
    Nuan Wen$^{2}$,
    Yanchen Liu$^{3,4}$,
    Po-Nien Kung$^{1}$\\
    {\bf Wenna Qin}$^{3}$,
    {\bf Shicheng Wen}$^{2}$,
    {\bf Azure Zhou}$^{3}$,
    {\bf Diyi Yang}$^{3}$,
    {\bf Xuezhe Ma}$^{2}$,
    {\bf Nanyun Peng}$^{1}$,
    {\bf Wei Wang}$^{1}$
}
\affiliations{
    $^{1}$University of California, Los Angeles
    \\
    $^{2}$University of Southern California
    \\
  $^{3}$Stanford University
  \\
    $^{4}$Harvard University
    \\
    \{ma, ataylor2, ponienkung, violetpeng, weiwang\}@cs.ucla.edu,
    \{nuanwen, wenshich, xuezhema\}@usc.edu
    \\
    yanchenliu@g.harvard.edu, \{wennaqin, amysz, diyiy\}@stanford.edu
}

\begin{document}

\maketitle

\begin{abstract}
    We present MIDDAG, an intuitive, interactive system that visualizes the information propagation paths on social media triggered by COVID-19-related news articles accompanied by comprehensive insights, including user/community susceptibility level, as well as events and popular opinions raised by the crowd while propagating the information. Besides discovering information flow patterns among users, we construct communities among users and develop the propagation forecasting capability, enabling tracing and understanding of how information is disseminated at a higher level. A demo video and more are available at \url{https://info-pathways.github.io}.
\end{abstract}

\section{Introduction}
\label{sec:intro}

The current information propagation ecosystem consisting of traditional news outlets and social media platforms allows for near real-time transmission of information concerning current events. The complexity of the multi-modal data contained in individual information pathway (IP) necessitates the development of novel approaches to analyze how information originally reported by traditional news outlets will spread across social media platforms, as well as the receptivity of the users engaging with this information. 
Many existing methods perform link prediction on social media networks~\cite{lp-survey-social-networks,lp-survey1,lp-survey2, lp-ego-structure} and IPs~\cite{lighting, PredictingIPkdd}. 
It is hard to find works that intuitively visualize the existing and forecasted IPs with comprehensive insights on motivating forces of information propagation, such as user/community characteristics and discussion content.

We present a comprehensive visualization system incorporating state-of-the-art information analysis components to provide a clear demonstration of information propagation details and patterns concerning COVID-19-related news within and across the audiences of prominent news organizations from countries highly impacted by COVID-19 across social media platforms. 
The first dataset consists of all COVID-19-related tweets extracted from the Twitter API. 
Using tweets containing links to news articles authored by selected news organizations, we construct communities of users based on their engagement patterns with the selected news organizations, which enables us to construct and visualize the IPs at the community level. 
We use these pathways to train our novel IP prediction model and present the predicted information propagation patterns for a held-out evaluation set. 
We also present a novel machine-learning-based approach trained on the collected tweets at the user level to predict and display the susceptibility level of users and communities, which provides further insight into each pathway. 
In addition to the susceptibility score, we perform event extraction on each tweet in an IP, which shows how and if the core ideas of the original article spread. 
To diversify the source of our data, we also use the Reddit data available for each news organization and apply existing techniques to extract popular opinions from the available posts in order to provide additional characterization to the IPs once visualized. Finally, we pack the IP analysis capabilities in a system for the user to select, visualize, interact and discover IPs at both user and community levels.

\section{Data and System Design}

\begin{figure*}
    \centering
    \includegraphics[width=0.88\textwidth]{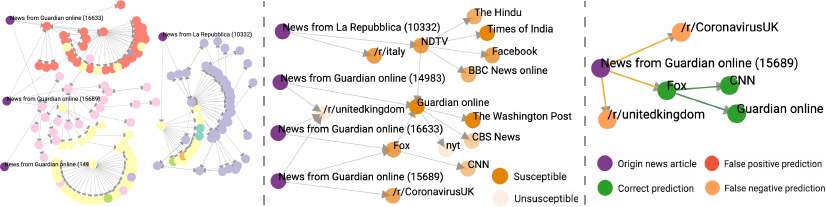}
    \caption{Visualizations of the user-level information pathways (left), community-level information pathways with estimated susceptibility level for each community (center), and community-level information propagation forecasting results (right).}
    \label{fig:demo_ui}
\end{figure*}

We use COVID-19-related social media data on Twitter and Reddit to enable a broad IP investigation.
We utilize an updated version of the Twitter news dataset presented in prior work \cite{lighting}, including all tweets from May 2020 to April 2021 containing COVID-19 keywords. Due to the large dataset size, we focus on the subset containing tweets from May 15 to May 30, 2020, which contained the largest number of tweets among all periods, including 640 million tweets from 5.3 million users. 
We consider news articles from the most prominent organizations (according to the Digital News Report from the Reuters Institute) for 15 countries with the most COVID-19 cases as the start of pathways. 
We retrieve all user-level IPs, including source tweets mentioning a news URL and subsequent retweets and replies.
Besides Twitter, we explored discussions on Reddit with hyperlinked COVID-19 news articles and within the same period as the Twitter subset. In total, we collected 5,410 posts on 4,578 unique articles across 649 subreddit communities.

Our system demonstrates IPs and their properties. The user first types keywords to search and picks news articles that serve as the starting points of IPs. The user-level visualization includes reply/repost propagation among users, susceptibility score and community assignment. Community-level visualization consists of a directed graph indicating IPs among communities, community aggregated susceptibility levels and key opinions. The event panel shows the event trigger, type and associated arguments in users' discussions.

\section{Components}

We first construct communities among social media users to aggregate the user-level IP to the community granularity. 
We develop ML models to forecast the information flow, and estimate the susceptibility levels. 
We further demonstrate events and leading opinions in users' discussions.

\mypar{Community Construction}
\label{community}
We assign users to communities centered around specific news organizations. 
The community aggregation method follows prior work that measures the influence of nodes in social networks \cite{InfluenceAndPassivity}. It measures the influence that a given node exerts on its neighborhood and the likelihood of how passive or receptive to propagation from its neighborhood a given node is.
For each user that interacts with the given community, we retrieve the number of URLs it has posted and compile its one-hop neighborhood based on the source tweets from the community in question it has both posted and interacted with; these individual user graphs are then united to construct a directed community graph. 
We then assign weights to each user where $Q_{i}$ is the number of URLs that $i$ mentioned and $S_{ij}$ is the number of URLs mentioned by $i$ and retweeted by $j$ as illustrated in the following equations. Finally, we perform the Influence-Passivity algorithm until the Influence ($I_i$) and Passivity ($P_i$) scores have converged. 
$$
\resizebox{0.47\textwidth}{!}{$
w_e = \frac{S_{ij}}{Q_{ij}}, 
u_{ij} =
\displaystyle \frac{w_{i,j}}{\displaystyle \sum_{k:(k,j)\in E}w_{kj}},
v_{ji} = \displaystyle \frac{1-w_{ji}}{\ \displaystyle \sum_{k:(j,k)\in
E}(1-w_{jk})};
I_i \leftarrow \displaystyle \sum_{j:(i,j)\in E}u_{ij}P_j,
\label{eqn:passivity_update}
P_i \leftarrow \displaystyle \sum_{j:(j,i)\in E}v_{ji}I_j
$
}
$$

\mypar{Pathway Prediction}
\label{pathway_prediction}

We use an improved version of the state-of-the-art IP prediction model~\cite{lighting}. 
We use the community assignments described above to map each user-level information pathway to the community level. 
The community-level pathways of time periods other than the 15-day selected duration are separated into distinct time windows for training and evaluation, and a graph neural network model is trained to perform link prediction. The evaluation results show that the IP prediction model yields 86.83\% AUC for the link prediction task. We apply the trained model to the unseen IP graphs to conduct autoregressive prediction of the full information propagation traces. 

\mypar{Susceptibility}
\label{sus_prediction}
Users' susceptibility reflects their reactions to a piece of misinformation. Since collecting an unobservable belief is hard, we develop an ML model to estimate susceptibility by analyzing its influence on users' repost behavior ($P_{repost}$)~\cite{Liu2023ScrollMisbeliefModeling}. When a user ($u$) perceives a piece of content ($c$), a more susceptible user is more likely to repost $c$. The reposting probability is calculated by 
$$
P_{repost} = Sigmoid (E(u) * E(c) * Sus(E(u), E(c)))
$$
where $u$ is represented by user history posts, $E$ is the text embedding model based on RoBERTa-large~\cite{reimers-2019-sentence-bert}, and the susceptibility score is produced by the $Sus$ - an MLP with the user and content embeddings as input. We perform contrastive learning to tune the model to distinguish reposted $(u, t)$ pairs from non-reposted ones. 
We train the ML model with misinformation tweets in the ANTi-Vax~\cite{HAYAWI202223} and CoAID~\cite{cui2020coaid} datasets and we retrieve corresponding user profiles through the Twitter API. The model produces 86.28\% F1 score for retweet behavior prediction, indirectly indicating its reliable performance for susceptibility modeling. To obtain an aggregated susceptibility score for a community, we calculate the mean of individual susceptibility scores.

\mypar{Event Extraction and Community Opinion}
\label{event_extraction_com_opinion}
Event extraction (EE) aims to identify triggers and arguments for events mentioned in the text, which enables us to understand users' discussion at a scale~\cite{Ma2021EventPlusTemporalEvent,Ma2023DICEDataEfficientClinical,ma-etal-2024-star}. We first define a COVID-19 related event ontology including 9 event types (\ie end organization, social distancing, lock down, quarantine, vaccinate, die, fine, transport and extradite) and their arguments. Since there are no existing EE annotations on COVID-related events, we perform instruction tuning to enable a T5-large model~\cite{Raffel2023ExploringLimitsTransfer} to generalize to newly defined event types following UIE~\cite{Lu2022UnifiedStructureGenerationa}
by pre-training it with text-structure pairs and further fine-tuning it on 13 datasets of entity/relation/event/sentiment extraction tasks encoded with a unified language. 
To provide discussion details among users in a community when it propagates certain information, we identify the most liked post among downstream posts in an information flow as the representative opinion of the community.

\section*{Acknowledgments}

Many thanks to the anonymous reviewers for their feedback. 
This effort was sponsored by the Defense Advanced Research Project Agency (DARPA) grant HR00112290103.

\bibliography{ma_auto,custom}

\end{document}